# Room-temperature ferromagnetism in epitaxial bilayer FeSb/SrTiO$_3$(001) terminated with a Kagome lattice

Huimin Zhang,[1,2]†* Qinxi Liu,[3]† Liangzi Deng,[4]† Yanjun Ma,[1] Samira Daneshmandi,[4] Cheng Cen,[1,5] Chenyu Zhang,[6] Paul M. Voyles,[6] Xue Jiang,[2,3]* Jijun Zhao,[2,3] Ching-Wu Chu,[4] Zheng Gai,[7] and Lian Li[1]*

[1]Department of Physics and Astronomy, West Virginia University, Morgantown, WV 26506, USA
[2]State Key Laboratory of Structural Analysis, Optimization and CAE Software for Industrial Equipment, Dalian University of Technology, Dalian, 116024, China
[3]Key Laboratory of Materials Modification by Laser, Ion and Electron Beams (Dalian University of Technology), Ministry of Education, Dalian 116024, China
[4]Department of Physics and Texas Center for Superconductivity, University of Houston, Houston, 77204, TX, USA
[5]Beijing National Laboratory for Condensed Matter Physics and Institute of Physics, Chinese Academy of Sciences, Beijing 100190, China
[6]Department of Materials Science and Engineering, University of Wisconsin–Madison, Madison, WI 53706, USA
[7]Center for Nanophase Materials Sciences, Oak Ridge National Laboratory, Oak Ridge, TN, 37831 USA

†These authors contributed equally: Huimin Zhang, Qinxi Liu, Liangzi Deng.
*Correspondence to: huiminzhang@dlut.edu.cn, jiangx@dlut.edu.cn, lian.li@mail.wvu.edu

**Abstract**

Two-dimensional (2D) magnets exhibit unique physical properties for potential applications in spintronics. To date, most 2D ferromagnets are obtained by mechanical exfoliation of bulk materials with van der Waals interlayer interactions, and the synthesis of single or few-layer 2D ferromagnets with strong interlayer coupling remains experimentally challenging. Here, we report the epitaxial growth of 2D non-van der Waals ferromagnetic bilayer FeSb on SrTiO$_3$(001) substrates stabilized by strong coupling to the substrate, which exhibits in-plane magnetic anisotropy and a Curie temperature above 300 K. *In-situ* low-temperature scanning tunneling microscopy/spectroscopy and density-functional theory calculations further reveal that a Fe Kagome layer terminates the bilayer FeSb. Our results open a new avenue for further exploring emergent quantum phenomena from the interplay of ferromagnetism and topology for application in spintronics.

Ferromagnetic (FM) order in two-dimensional (2D) materials has attracted extensive attention owing to its tremendous application prospects in data storage[1], sensors[2], high-efficiency spin-based computers, and other nanoscale devices[3,4]. Understanding emergent phenomena in 2D hybrid magnetic structures is crucial in laying a solid foundation for realizing these potential applications[5,6]. To date, substantial theoretical[7] and experimental[8] efforts have been made to achieve long-range ferromagnetic order in 2D materials at room temperature (RT). Such order was initially predicted to be absent at finite temperatures ($T > 0$) for isotropic 2D systems based on the Mermin-Wagner theorem[9]. However, magnetic anisotropy was later found to quench thermal fluctuations[10,11,1] and help stabilize the long-range FM order in the 2D limit. Most recently discovered 2D magnets are based on layered materials[11–17], in which the layers are held together by van der Waals (vdW) interactions. Even though 2D magnetism is well established in the single-layer limit, the FM orders in the vdW materials are generally fragile, with Curie temperature ($T_c$) well below RT. For instance, the $T_c$ in $CrI_3$[12], $CrCl_3$[18], $CrBr_3$[19], $Cr_2Ge_2Te_6$[11], and CrSBr[20] single-layers are reported to be 45 K, 13 K, 27 K, 44 K, and 146 K, respectively. This has been attributed to enhanced spin fluctuations at the reduced dimensions or relatively weak exchange interactions. In addition, 2D vdW materials are usually not air-stable, and the size of flakes exfoliated from bulk crystals is typically limited to micrometers, which is nonideal for practical applications.

One route to overcome these drawbacks is to develop 2D non-vdW magnetic materials[21] that exhibit different magnetic properties than their bulk counterparts. For example, 2D hematene (α-$Fe_2O_3$) is a ferromagnet, while bulk hematite is antiferromagnetic[22]. These materials have already shown unique advantages of easily modulated spin behavior and high Curie temperature[21], as observed in hematene (α-$Fe_2O_3$)[22], ilmenene ($FeTiO_3$)[23], chromiteen ($FeCr_2O_4$)[24], and transition-metal chalcogenides (TMDCs) including $Cr_2S_3$[25], $CrSe$[26], $CrTe$[27], $Cr_2Te_3$[28], $MnTe$[29], and *h*-$FeTe$[30].

Another non-vdW material class is Kagome magnets, which exhibit emerging topological behavior with a complex magnetic phase diagram due to frustration and competing magnetic interactions[31]. For example, FeSn, consisting of Sn honeycomb layers and corner-sharing $Fe_3Sn$ Kagome ferromagnetically coupled in-plane and antiferromagnetically out-of-plane, hosts linearly dispersing Dirac states and flat bands[32]. Exploring these 2D non-vdW materials with long-range ferromagnetic order is an active area of research to probe the interplay of geometrical frustration, topology, and magnetism.

In this work, we synthesize bilayer FeSb films on $SrTiO_3$(001) substrates by molecular beam epitaxy (MBE) and perform *in-situ* low-temperature scanning tunneling microscopy/spectroscopy (LT-STM/S), *ex-situ* scanning transmission electron microscopy (STEM), magnetization measurements, and density-functional theory (DFT) calculations. The results show that while bulk FeSb is an antiferromagnet[33], the bilayer FeSb film is a non-vdW ferromagnet stabilized by strong coupling to the STO substrate, which exhibits a robust long-range FM order above 300 K with in-plane anisotropy. Furthermore, LT-STM/S measurements and DFT calculations show that the bilayer FeSb is terminated by a surface Fe Kagome lattice, offering a unique 2D system exploring emergent quantum phenomena from the interplay of ferromagnetism and topology.

**RESULTS AND DISCUSSION**

**Epitaxial growth of FeSb/SrTiO$_3$(001) films. Figure 1a** presents a typical topographic STM image of FeSb films grown on Nb:SrTiO$_3$ (STO) (001) substrates (see Methods for growth conditions). The surface morphology of the FeSb film is conformal to the step-terrace topography of the STO substrate. The growth of FeSb film follows a layer-by-layer growth mode (Fig. S1). The thickness of the FeSb film is determined to be ~1.0 nm based on the analysis of line profiles as shown in Fig. 1c, consistent with a bilayer FeSb (Supplementary Note 1, space group $P6_3/mmc$ with the lattice constant $a$ = 0.4065 nm and $c$ = 0.5121 nm). This is further confirmed by scanning transmission electron microscopy (STEM) images (Supplementary Note 2 and Fig. S2), where the interface between FeSb and the $TiO_2$-terminated STO is resolved with an epitaxial relationship of [11-20]FeSb//[100]STO. In particular, the two nearby Sb atom rows are clearly resolved with an Sb-Sb spacing of 4.1 Å, consistent with the theoretical value of 4.06 Å for FeSb. No discernible defects or extra Fe atoms in the film are observed in the STEM images.

The 2nd bilayer starts to grow only after the 1st bilayer film fully covers the STO substrate (Fig. S1). We notice that the FeSb films exhibit different thermal stability as the 2nd bilayer starts to desorb when the substrate temperature increases to 438 °C (Fig. S3). Moreover, thicker films beyond the 2nd bilayer exhibit a different structural phase (Fig. S4). These observations suggest that the FeSb film is an interface-stabilized phase with a strong coupling between the 1st bilayer and the STO substrate, which is supported by DFT calculations to be discussed below.

The surface topography of the FeSb film exhibits multiple domains tens of nanometers (see STM images in Figs. 1b, S5, and S6), likely due to the growth of FeSb on STO(001) substrate is symmetry mismatched with a three-fold symmetric material epitaxially grown on the four-fold symmetric STO (001) substrate (Figs. S7 and S8). Consequently, a polydomain film is formed by multi-grain coalescence during the growth [34,35].

Within each domain, atomic resolution STM imaging reveals a Kagome lattice composed of hexagons and shared triangles denoted by the dashed pattern (Figs. 1e-f, S5). The periodicity of the Kagome lattice is 1.7 nm, four times the lattice constant $a_{FeSb}$ = 4.01 Å. Within the Kagome lattice, higher contrast is observed at the center and lower contrast at the sharing triangle. From the line profile shown in Fig. 1d, the high/low contrast gives rise to a difference of ~1.2 Å, which exhibits minimum change under various bias $V$ ranging from 1.0 V to -0.2 V (Fig. S9). The Kagome structure observed here is similar to the previously reported superstructure formed by Fe deposited on the Sb(111) surface [36].

**Electronic properties of the Kagome lattice.** Multiple peaks are observed in the dI/dV spectrum (Fig. 1g), and the nonzero density of states (DOS) at the Fermi level ($E_F$) indicates a metallic behavior [37,38], which is also site-dependent (Fig. S10). For the STM image in Fig. 2a, dI/dV were measured at three typical sites A, B, and C: the center of the hexagon, the center, and the corner of the shared triangle site (Fig. 2b). Three pronounced peaks are marked by red (-0.66 eV), black (0.18 eV), and purple (0.69 eV) arrows in Fig. 2b, indicating higher DOS from the shared triangle site (B or C site) at these energies. This is directly reflected in the dI/dV maps shown in Figs. 2c-h (and more data in Fig. S11). The DOS is higher at the shared triangle corners, and the hexagon's center for the map g(r, -0.9 eV) (Fig. 2c). In Fig. 2d, the map $g(\mathbf{r}, -0.7\text{ eV})$ highlights the shared triangle site. It was previously reported that higher DOS is observed at the shared triangle sites near the energy of the flat band in dI/dV maps of a Kagome lattice from twisted silicene multilayer [39]. Similarly, the peak at ~-0.7 eV marked by the red arrow in Fig. 2b likely corresponds to the flat band of the Kagome lattice on the surface of BL FeSb/STO. This is consistent with the map $g(\mathbf{r}, -0.7\text{ eV})$, where the center of the Kagome lattice exhibits low contrast, whereas the shared triangle sites show higher contrast. For energies closer to the Fermi level, the map $g(\mathbf{r}, -0.6\text{ eV})$ (Fig. 2e) and $g(\mathbf{r}, -0.2\text{ eV})$ (Fig. 2f) show structural details of the Kagome lattice's center site, either a ring structure or a six-lobe feature. In contrast, above the Fermi level, the map $g(\mathbf{r}, 0.2\text{ eV})$ (Fig. 2g) has higher DOS at the center of the shared triangle site. In the map $g(\mathbf{r}, 0.8\text{ eV})$ (Fig. 2h), the center site of the Kagome lattice exhibits higher contrast. The appearance of the Kagome lattice is reproduced by DFT calculations, which will be discussed below.

**Magnetization properties of 1BL FeSb/STO films.** We performed *ex-situ* magnetization measurements for FeSb/STO films after capping with ~ 20 nm amorphous Sb. The results for two samples with 60% and 100% coverages (Figs. 3a&b) are displayed in Figs. 3c&d. Clear hysteresis is observed in the *M-H* plots. As the temperature increases from 4.2 K, 150 K to 300 K, the hysteresis loops shrink with fields ($H$), characteristic of ferromagnetic order. The *M-H* loops at 300 K are apparent, indicating a room-temperature ferromagnetism, consistent with temperature-dependent *M* vs. *T* measurements (See Fig. S12 for the *M-T* plots). We also noticed a diamagnetic background in the raw data of the FeSb/STO films, which was subtracted by corresponding linear fittings (Fig. S13). Moreover, the FM signal is confirmed by the magnetization characterization of three additional 1.0 BL samples (Fig. S14), where *M-H* loops are observable even at $T = 390$ K, indicating a Curie temperature $T_C > 390$ K. The external magnetic field, $H$, is applied parallel to the sample plane during the above measurements. In contrast, under out-of-plane magnetic fields, there is a minor change in the *M-H* loops (see Fig. S15). Therefore, we conclude that the easy axis aligns along the in-plane direction. As the coverage increases from 60% to 100%, the coercive field $H_c$ decreases from 485.3 Oe to 374.7 Oe at $T = 4.2$ K, likely due to the change of the film from isolated islands (single magnetic domains) to percolated from 60% coverage to full bilayer (Fig. S6). The saturation magnetization $M_s$ for 60% coverage is $\sim 5\times 10^{-6}$ emu (Fig. 3c), comparable to the full bilayer film $\sim 1.0\times 10^{-5}$ emu (Fig. 3d), respectively.

We have carried out control experiments to rule out extrinsic effects that could be responsible for the robust ferromagnetism presented above. First, the ~20 nm Sb-capped STO substrate only shows a weak ferromagnetic order below 100 K and a diamagnetic order above 200 K (see Fig. S16), consistent with that observed in SrTiO$_3$ single crystal[40]. Consequently, the ferromagnetic signal observed here originates from the epitaxial 1BL FeSb/STO films rather than the Sb capping layer or the STO substrate. In addition, the STO substrate only contributes to the diamagnetic background in the raw data of FeSb/STO films (Fig. S13).

Next, we determine if excess Fe atoms in the film could contribute to the observed ferromagnetic order. This possibility can be ruled out based on the following two observations. First, no substantial extra Fe atoms or clusters are observed by STM (Fig. S1) or STEM imaging (Fig. S2). Second, we deposited Sb atoms onto the surface of the FeSb films and did not observe the formation of extra FeSb islands (Fig. S17). If there were excess Fe atoms in the film, they might react with Sb and form additional FeSb islands, similar to the case of FeSe/STO[41]. This is also confirmed by the fact that the $M_s$ is similar for FeSb films with coverages of 60% and 100%, because if the interstitial Fe atoms play a role, then the full bilayer film should exhibit higher $M_s$ than the film with 60% coverage. Based on the above, we conclude that ferromagnetism originates from the FeSb/STO films.

**Density-functional theory calculations**. We further performed DFT calculations to determine the structure of FeSb films. Among all the six phases (Supplementary Note 1), our simulations indicate that only the ultrathin films with FeSb phase (space group P6$_3$/mmc with the lattice constant $a$ = 0.4065 nm and $c$ = 0.5121 nm)[41,42] are energetically favorable on TiO$_2$-terminated SrTiO$_3$(001) substrate. The thickness of the film ~1.0 nm corresponds to bilayer FeSb (Fig. S18). The epitaxial relationship between the FeSb and the TiO$_2$-terminated STO substrate, their lattice mismatch, and the orientation of the FeSb polycrystals are shown in Supplementary Note 3. We constructed the structure model based on DFT calculations and detailed analysis of the bias-dependent STM images. First, the topmost bright protrusions in the STM images consist of six bright features arranged in a regular hexagonal configuration, as shown in Figs. S5 and S19c. Next, we calculated the adsorption energy ($E_{ab}$) of Sb adatom on different sites of a bilayer FeSb, defined as $E_{ab} = E_{total} - E_{film}^{FeSb} - E_{atom}^{Sb}$, where $E_{total}$ and $E_{film}^{FeSb}$ are the energies of the structure configuration with and without absorbed Sb atoms, and $E_{atom}^{Sb}$ is the energy of single Sb atom. Figures S19a&b show four typical sites labeled 1, 2, 3, and 4, and their corresponding adsorption energy ($E_{ab}$) of Sb adatom. Negative adsorption energy is observed for sites 1 and 2, with the lowest adsorption energy present at site 2 (-1.73 eV/Å$^2$), indicating that the energy will decrease with the Sb atom absorbed. We construct a surface adsorption structural model as shown in Fig.

4a with site 2 adsorbed and find the corresponding simulated STM image (Fig. 4b) perfectly fit the experimental one (Fig. 4c). Both the enhanced DOS at the center (purple hexagon) and the suppressed DOS at the hollow site (cyan triangle) are reproduced. The calculated interface interaction energy $E_{inter}$ of FeSb films exhibits a minimum for the bilayer (Supplementary Note 5 and Fig. S18e), suggesting it is energetically stable than the ML and TL, consistent with the bilayer growth observed by STM. Note that since FeSb is stacked alternatively by Fe and Sb atomic layers along the *c*-direction, the film can either be Fe-terminated or Sb-terminated in principle (Supplementary Note 6). However, the reconstructed Sb-termination is triangular rather than hexagonal (Fig. S20), which conflicts with the observation that STM imaging is independent of energy (Fig. S9). Thus, the possibility of Sb termination is ruled out.

Overall, the adsorption of Sb atoms on the Fe-termination leads to the modulated electronic structure, which contributes to the Kagome lattice observed by STM, i.e., the Kagome layer is an Sb-induced reconstruction of the surface Fe atoms. This structure can host quantum states arising from the interplay of spin-orbit coupling, electron correlations, and topology[43,44]. Our preliminary investigation has shown evidence of edge states (Fig. S21), suggesting topological order in the BL FeSb/STO films.

**Insight into the origin of ferromagnetic order in 1BL FeSb/STO films.** Based on DFT calculations, we discuss the intrinsic magnetism of FeSb films (Supplementary Note 7). In FeSb, each Fe atom is bonded to six equivalent Sb atoms to form $FeSb_6$ octahedra, and the Fe-3*d* orbitals split due to trigonal antiprismatic crystal field (Fig. 4e). FeSb films exhibit long-range magnetic ordering and the total magnetic moment *M* is contributed by Fe atoms. The calculated *M* is 2.04 $\mu_B$ per Fe for bilayer FeSb (Table S2), moderately smaller than the experimental estimation of 2.6 $\mu_B$ per Fe atom. The calculated magnetic anisotropy energy (MAE) is 0.412 meV per BL (Table S3). Interestingly, it favors an in-plane anisotropy, consistent with the experimental observation. The bilayer FeSb is a metallic system evidenced by the calculated DOS (Fig. S22). The calculated value of $D(E_F) \times I$ is 3.17, which meets the Stoner criterion, i.e., $D(E_F) \times I > 1$ (Table S5), signifying itinerant ferromagnetism. Based on the mean-field Ising model, the Curie temperature $T_C$ is calculated to be is 377.9 K (Supplementary Note 7), consistent with our experimental results (> 390 K).

**Interfacial effect in 1BL FeSb/STO films.** Figure 4d shows interfacial charge densities across the FeSb film (top) and the STO substrate (bottom) from DFT calculations. Separation of blue (electron accumulation) and yellow (electron depletion) regions is revealed, indicating charge transfer between the STO and FeSb films. The charge transfer suggests a strong interaction between the FeSb films and the STO substrates. To address whether the accumulated charge is responsible for the emergent ferromagnetic order in FeSb/STO films, we grew FeSb films on double-layer graphene/SiC(0001) substrates, where a minimum interfacial effect is expected (Fig. S23)[45,46]. Similar magnetic properties, including hysteresis loops persisting at T = 300 K, suggest that the ferromagnetic order is an intrinsic feature of the FeSb films rather than an interfacial effect. Moreover, we notice a slightly smaller $H_c$ and $M_s$ for FeSb films grown on the graphene/SiC(0001), which could be attributed to the more randomly distributed FeSb islands from the Volmer-Weber growth mode (Fig. S23) in contrast to the layer-by-layer growth on the STO substrate (Fig. S1). The change of growth mode also confirms the much stronger interfacial interaction between the 1BL FeSb/STO interface.

**CONCLUSIONS**

In summary, we successfully grew bilayer FeSb films on $TiO_2$-terminated $SrTiO_3$(001) substrates by MBE, an interface-stabilized 2D FM magnet terminated by a Fe Kagome layer. The films exhibit a ferromagnetic $T_c$ > 300K and an in-plane magnetic anisotropy, offering opportunities for further exploring emergent quantum phenomena from the interplay of ferromagnetism and topology with potential applications in spintronics.

**METHODS**

*Sample preparation.* The FeSb films were prepared by MBE on Nb-doped (0.05 wt.%) $SrTiO_3$(001) substrates. To achieve a flat surface with step-terrace morphology, the STO substrates were first degassed at 873 K for 3 h, followed by annealing at 1273 K for 1 h. High purity Fe (99.995%) and Sb (99.9999%) were evaporated from a home-made Tantalum boat and a K-cell onto the Nb-doped $SrTiO_3$(001) substrate at temperatures between 573-623 K during growth at a growth rate of 0.1 BL/min. During growth, the Sb/Fe ratio is estimated to be 5:1. The as-grown FeSb films were capped with ~20 nm amorphous Sb films before being transferred out of the vacuum for magnetization measurements.

*Low-temperature STM/STS.* Low-temperature STM/S measurements were carried out using a polycrystalline PtIr tip in a Unisoku-1300 STM system. The samples were transferred *in-situ* to the STM chamber. During STM measurements, the sample was kept at $T = 4.67$ K with PtIr tips calibrated on Ag/Si(111) films. The STS was acquired using a standard lock-in technique with a bias modulation $V_{rms} = 20$ mV at 732 Hz unless otherwise specified.

***Magnetization measurement.*** The magnetic properties of the samples were measured in a Magnetic Property Measurement System (MPMSXL7) by Quantum Design with temperatures up to 390 K and magnetic fields up to 1000 Oe, and a Physical Property Measurement System (PPMS) with a VSM option by Quantum Design with a sensitivity of ~$10^{-8}$ emu.

***Density-functional theory calculations.*** First-principles calculations were carried out with VASP. Spin-polarized generalized gradient approximation with Perdew-Burke-Enzerhof parameterization was used as the exchange-correlation functional, combined with projector-augmented wave pseudopotentials as well as a plane wave basis set with energy cutoff at 500 eV[47]. Correction of van der Waals interactions using the DFT-D3 scheme was included in all calculations[48]. A vacuum space of 20 Å thickness was added to avoid interaction between FeSb slabs. During the calculations, a Monkhorst–Pack $k$-point mesh of 0.02 Å$^{-1}$ was chosen for sampling the 2D Brillouin zones.

**Figures**

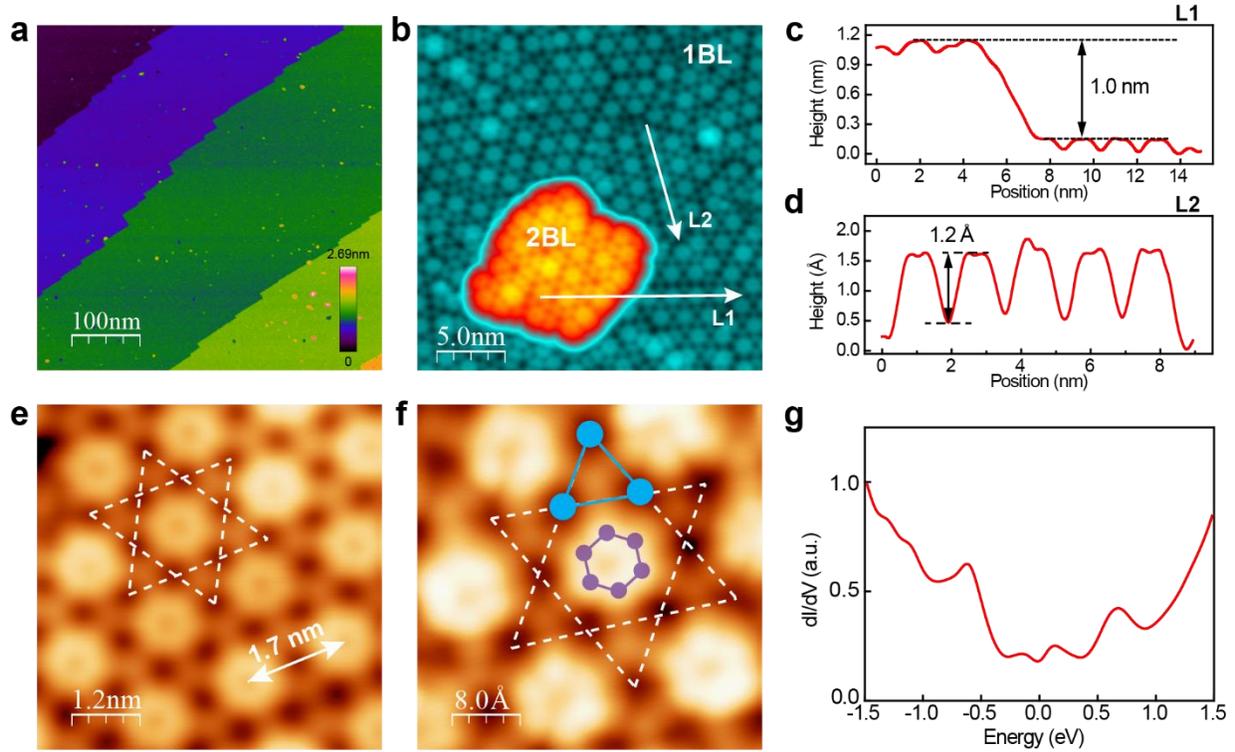

**Figure 1. MBE-grown FeSb films on SrTiO$_3$(001) substrate.** (a) Topographic STM image of 1 BL FeSb film epitaxially grown on the STO substrate. The step terraces in the image are from the STO substrate. Setpoint: $V$ = 2.0 V, $I$ = 10 pA. (b) Topographic STM image of a 2BL FeSb island. Setpoint: $V$ = 1.0 V, $I$ = 50 pA. (c) Line profile along the white arrow L1 in (b). The height of the 1BL film is determined to be 1.0 nm. (d) Line profile along the white arrow labeled L2 in (b). The high/low contrast of the surface reconstruction is 1.2 Å. (e) Atomic resolution STM image of the 1 BL FeSb film showing a Kagome lattice. Setpoint: $V$ = -0.1 V, $I$ = 600 pA. (f) Close-up view of the Kagome lattice revealing the enhanced density of state at the center (purple hexagon) and suppressed density of state at the shared triangle site (cyan triangle). Setpoint: $V$ = 0.1 mV, $I$ = 200 pA. (g) Typical dI/dV spectrum of the 1 BL FeSb film taken at the corner of the shared triangle site (cyan dot site in (f)).

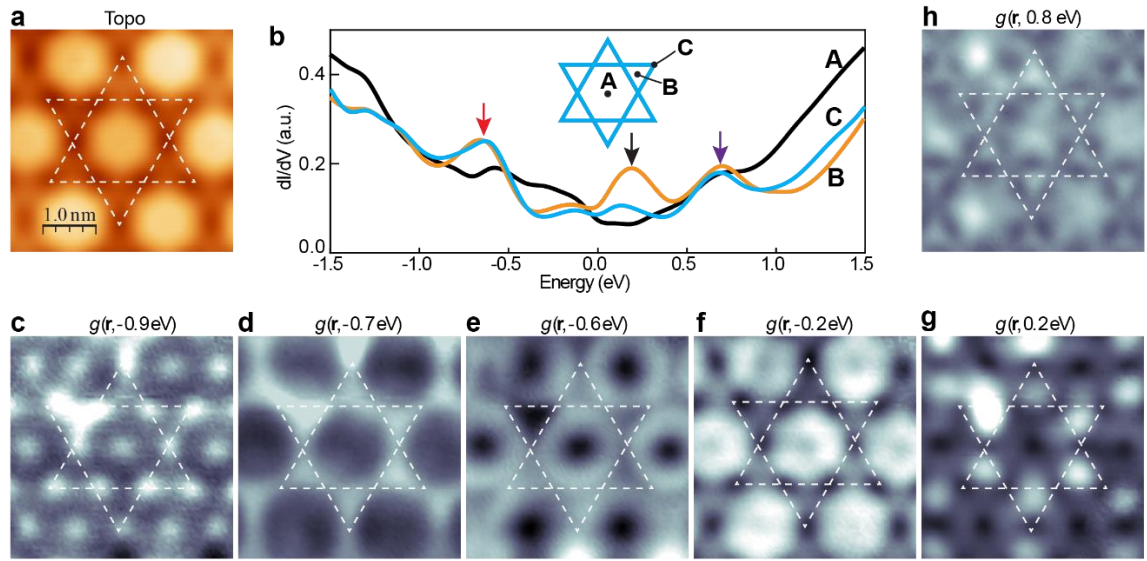

**Figure 2. dI/dV maps of the Kagome lattice on the surface of the 1BL FeSb/STO films.** (a) Topographic STM image of the Kagome lattice. Setpoint: $V = 1.0$ V, $I = 1.0$ nA. (b) dI/dV spectra taken at A (center), B (center of the shared triangle), and C (corner of the shared triangle) sites. The red, black, and purple arrows mark the peaks at -0.66 eV, 0.18 eV, and 0.69 eV, respectively. (c)-(h) Differential conductance maps at the energy specified. Setpoint: $V = 1.0$ V, $I = 1.0$ nA, and $V_{mod} = 20$ meV.

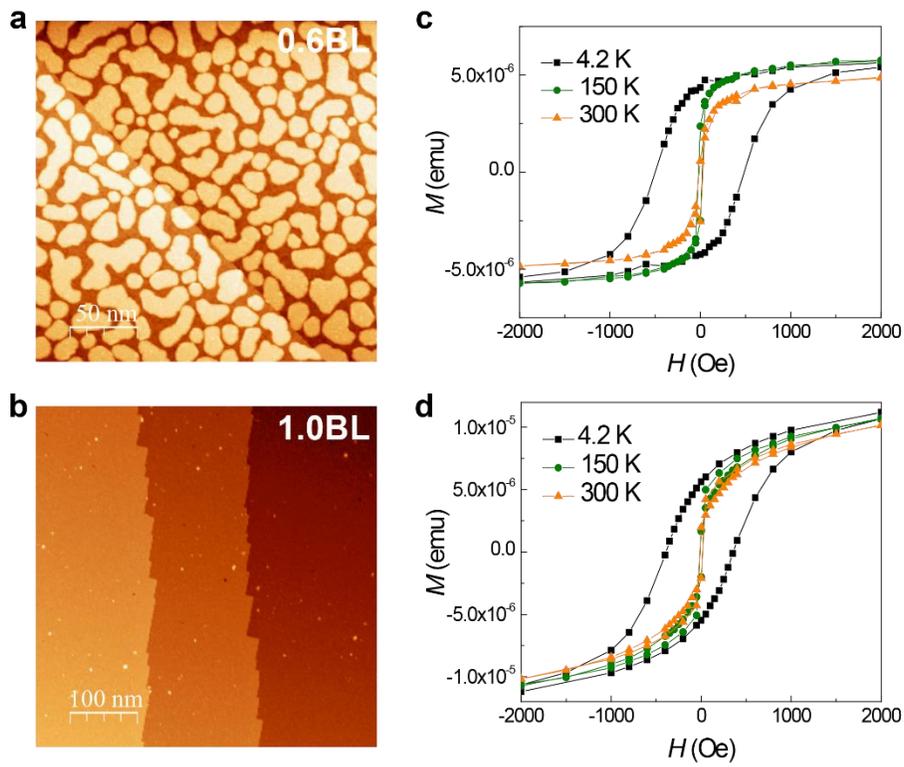

**Figure 3. Magnetic properties of FeSb/STO(001) films.** (a)-(b) Topographic STM images of FeSb films with coverages of 60% and 100%, setpoint: $V = 3.0$ V, $I = 20$ pA. (c)-(d) In-plane $M$-$H$ taken at 4.2 K, 150 K, and 300 K for the two films (The magnetization saturates at $H = 5000$ Oe, see Fig. S13 for details).

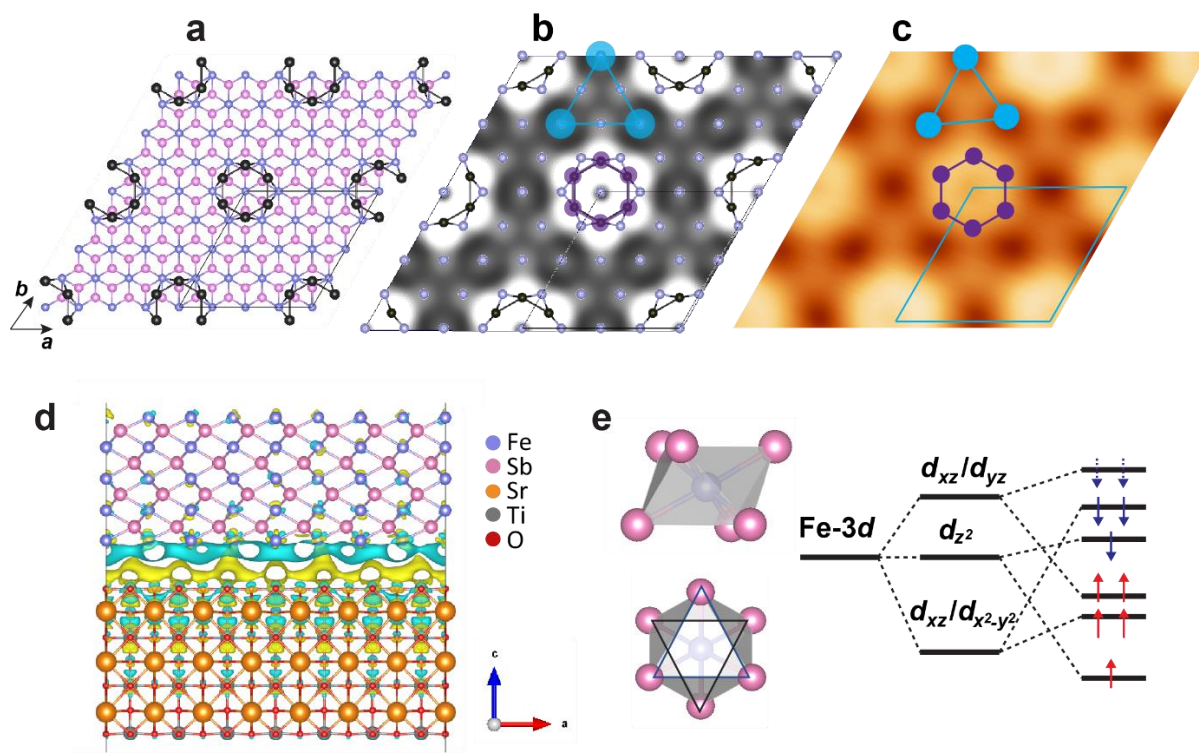

**Figure 4. Density-functional theory calculations of the surface, interface, and magnetism in 1BL FeSb/STO(001) films.** (a) Ball-and-stick model of the Fe terminated bilayer FeSb film with Sb reconstruction (black atoms on top). (b) Simulated STM image of the configuration in (a) at $E = 1.0$ eV with the surface Fe and adsorbed Sb atoms marked. The shared triangle sites of the Kagome lattice are exactly at the Fe site, as denoted by the cyan triangle. (c) Experimental STM image showing enhanced DOS at the protrusion (hexagonal purple ring) and suppressed DOS at the hollow site (cyan triangle). Setpoint: $V = -0.1$ V, $I = 600$ pA. (d) Charge density of FeSb/STO heterostructure. The isovalue is 0.0005 e Å$^{-3}$. Fe, Sb, Sr, Ti, and O atoms are denoted by blue, pink, orange, gray, and red balls. The blue and yellow colors represent charge depletion and accumulation, respectively. (e) Schematic diagram of coordinated FeSb$_6$ octahedron and the splitting of Fe-$3d$ orbitals under trigonal antiprismatic crystal field.

**ASSOCIATED CONTENT**
**Supporting Information**
**AUTHOR INFORMATION**

**Corresponding Author**

**Huimin Zhang** − State Key Laboratory of Structural Analysis, Optimization and CAE Software for Industrial Equipment, Dalian University of Technology, Dalian, 116024, China

Department of Physics and Astronomy, West Virginia University, Morgantown, West Virginia 26506, United States

Phone: (+86) 18010090159; Email: huiminzhang@dlut.edu.cn

**Xue Jiang** − State Key Laboratory of Structural Analysis, Optimization and CAE Software for Industrial Equipment, Dalian University of Technology, Dalian, 116024, China

Key Laboratory of Materials Modification by Laser, Ion and Electron Beams (Dalian University of Technology), Ministry of Education, Dalian 116024, China

Phone: (+86) 13940888317; Email: jiangx@dlut.edu.cn

**Lian Li** − Department of Physics and Astronomy, West Virginia University, Morgantown, West Virginia 26506, United States

Phone: (+1) 304-293-4270; Email: lian.li@mail.wvu.edu

**Authors**

**Huimin Zhang** − State Key Laboratory of Structural Analysis, Optimization and CAE Software for Industrial Equipment, Dalian University of Technology, Dalian, 116024, China

Department of Physics and Astronomy, West Virginia University, Morgantown, West Virginia 26506, United States

**Qinxi Liu** − Key Laboratory of Materials Modification by Laser, Ion and Electron Beams (Dalian University of Technology), Ministry of Education, Dalian 116024, China

**Liangzi Deng** − Department of Physics and Texas Center for Superconductivity, University of Houston, Houston, 77204, TX, USA

**Yanjun Ma** − Department of Physics and Astronomy, West Virginia University, Morgantown, WV 26506, USA

**Samira Daneshmandi** − Department of Physics and Texas Center for Superconductivity, University of Houston, Houston, 77204, TX, USA

**Cheng Cen** − Department of Physics and Astronomy, West Virginia University, Morgantown, WV 26506, USA

Beijing National Laboratory for Condensed Matter Physics and Institute of Physics, Chinese Academy of Sciences, Beijing 100190, China

**Chenyu Zhang** − Department of Materials Science and Engineering, University of Wisconsin–Madison, Madison, WI 53706, USA

**Paul M. Voyles** − Department of Materials Science and Engineering, University of Wisconsin–Madison, Madison, WI 53706, USA

**Xue Jiang** − State Key Laboratory of Structural Analysis, Optimization and CAE Software for Industrial Equipment, Dalian University of Technology, Dalian, 116024, China

Key Laboratory of Materials Modification by Laser, Ion and Electron Beams (Dalian University of Technology), Ministry of Education, Dalian 116024, China

**Jijun Zhao** − State Key Laboratory of Structural Analysis, Optimization and CAE Software for Industrial Equipment, Dalian University of Technology, Dalian, 116024, China

Key Laboratory of Materials Modification by Laser, Ion and Electron Beams (Dalian University of Technology), Ministry of Education, Dalian 116024, China

**Ching-Wu Chu** − Department of Physics and Texas Center for Superconductivity, University of Houston, Houston, 77204, TX, USA

**Lian Li** − Department of Physics and Astronomy, West Virginia University, Morgantown, WV 26506, USA

**Author contributions**

H.Z. and L.L. conceived and organized the study. H.Z. performed the MBE growth and STM/S measurements. L.D., Y.M, S.D., C.C., and Z.G. carried out magnetization characterizations. H.Z. and L.L. analyzed the data and wrote the paper. Q.L., X.J., and J.Z. performed DFT calculations. All authors discussed the results and commented on the paper.

**Notes**
The authors declare no competing financial interest.

**ACKNOWLEDGMENTS**

The MBE and STM work were supported by the U.S. Department of Energy, Office of Basic Energy Sciences, Division of Materials Sciences and Engineering under Award No. DE-SC0017632 and the National Science Foundation (Grant No. EFMA-1741673). Ching-Wu Chu acknowledges support from US Air Force Office of Scientific Research Grants FA9550-15-1-0236 and FA9550-20-1-0068, the T.L.L. Temple Foundation, the John J. and Rebecca Moores Endowment, and the State of Texas through the Texas Center for Superconductivity at the University of Houston (TcSUH). Huimin Zhang, Xue Jiang and Jijun Zhao acknowledge support from the National Natural Science Foundation of China (Grant Nos. 12304210, 11874097, 12274050, 91961204) and the Fundamental Research Funds for the Central Universities (DUT22LAB104, DUT22ZD103). The TEM work was supported by the US Department of Energy, Basic Energy Sciences (DE-FG02-08ER46547) and user facilities supported by the Wisconsin MRSEC (DMR-1720415).

## REFERENCES

(1) Song, T.; Cai, X.; Tu, M. W.-Y.; Zhang, X.; Huang, B.; Wilson, N. P.; Seyler, K. L.; Zhu, L.; Taniguchi, T.; Watanabe, K.; McGuire, M. A.; Cobden, D. H.; Xiao, D.; Yao, W.; Xu, X. Giant Tunneling Magnetoresistance in Spin-Filter van Der Waals Heterostructures. *Science* **2018**, *360* (6394), 1214–1218. https://doi.org/10.1126/science.aar4851.

(2) Zhong, D.; Seyler, K. L.; Linpeng, X.; Wilson, N. P.; Taniguchi, T.; Watanabe, K.; McGuire, M. A.; Fu, K.-M. C.; Xiao, D.; Yao, W.; Xu, X. Layer-Resolved Magnetic Proximity Effect in van Der Waals Heterostructures. *Nat. Nanotechnol.* **2020**, *15* (3), 187–191. https://doi.org/10.1038/s41565-019-0629-1.

(3) Soumyanarayanan, A.; Reyren, N.; Fert, A.; Panagopoulos, C. Emergent Phenomena Induced by Spin–Orbit Coupling at Surfaces and Interfaces. *Nature* **2016**, *539* (7630), 509–517. https://doi.org/10.1038/nature19820.

(4) Han, W.; Kawakami, R. K.; Gmitra, M.; Fabian, J. Graphene Spintronics. *Nat. Nanotechnol.* **2014**, *9* (10), 794–807. https://doi.org/10.1038/nnano.2014.214.

(5) Kezilebieke, S.; Huda, M. N.; Vaňo, V.; Aapro, M.; Ganguli, S. C.; Silveira, O. J.; Głodzik, S.; Foster, A. S.; Ojanen, T.; Liljeroth, P. Topological Superconductivity in a van Der Waals Heterostructure. *Nature* **2020**, *588* (7838), 424–428. https://doi.org/10.1038/s41586-020-2989-y.

(6) Yilmaz, T.; Tong, X.; Dai, Z.; Sadowski, J. T.; Schwier, E. F.; Shimada, K.; Hwang, S.; Kisslinger, K.; Kaznatcheev, K.; Vescovo, E.; Sinkovic, B. Emergent Flat Band Electronic Structure in a VSe2/Bi2Se3 Heterostructure. *Commun Mater* **2021**, *2* (1), 1–8. https://doi.org/10.1038/s43246-020-00115-w.

(7) Jiang, X.; Liu, Q.; Xing, J.; Liu, N.; Guo, Y.; Liu, Z.; Zhao, J. Recent Progress on 2D Magnets: Fundamental Mechanism, Structural Design and Modification. *Applied Physics Reviews* **2021**, *8* (3), 031305. https://doi.org/10.1063/5.0039979.

(8) Xu, H.; Xu, S.; Xu, X.; Zhuang, J.; Hao, W.; Du, Y. Recent Advances in Two-Dimensional van Der Waals Magnets. *Microstructures* **2022**, *2* (2), 2022011. https://doi.org/10.20517/microstructures.2022.02.

(9) Mermin, N. D.; Wagner, H. Absence of Ferromagnetism or Antiferromagnetism in One- or Two-Dimensional Isotropic Heisenberg Models. *Phys. Rev. Lett.* **1966**, *17* (22), 1133–1136. https://doi.org/10.1103/PhysRevLett.17.1133.

(10) Huang, B.; Clark, G.; Navarro-Moratalla, E.; Klein, D. R.; Cheng, R.; Seyler, K. L.; Zhong, D.; Schmidgall, E.; McGuire, M. A.; Cobden, D. H.; Yao, W.; Xiao, D.; Jarillo-Herrero, P.; Xu, X. Layer-Dependent Ferromagnetism in a van Der Waals Crystal down to the Monolayer Limit. *Nature* **2017**, *546* (7657), 270–273. https://doi.org/10.1038/nature22391.

(11) Gong, C.; Li, L.; Li, Z.; Ji, H.; Stern, A.; Xia, Y.; Cao, T.; Bao, W.; Wang, C.; Wang, Y.; Qiu, Z. Q.; Cava, R. J.; Louie, S. G.; Xia, J.; Zhang, X. Discovery of Intrinsic Ferromagnetism in Two-Dimensional van Der Waals Crystals. *Nature* **2017**, *546* (7657), 265–269. https://doi.org/10.1038/nature22060.

(12) Huang, B.; Clark, G.; Navarro-Moratalla, E.; Klein, D. R.; Cheng, R.; Seyler, K. L.; Zhong, D.; Schmidgall, E.; McGuire, M. A.; Cobden, D. H.; Yao, W.; Xiao, D.; Jarillo-Herrero, P.; Xu, X. Layer-Dependent Ferromagnetism in a van Der Waals Crystal down to the Monolayer Limit. *Nature* **2017**, *546* (7657), 270–273. https://doi.org/10.1038/nature22391.

(13) Liu, W.; Dai, Y.; Yang, Y.-E.; Fan, J.; Pi, L.; Zhang, L.; Zhang, Y. Critical Behavior of the Single-Crystalline van Der Waals Bonded Ferromagnet Cr2Ge2Te6. *Phys. Rev. B* **2018**, *98* (21), 214420. https://doi.org/10.1103/PhysRevB.98.214420.

(14) Deng, Y.; Yu, Y.; Song, Y.; Zhang, J.; Wang, N. Z.; Sun, Z.; Yi, Y.; Wu, Y. Z.; Wu, S.; Zhu, J.; Wang, J.; Chen, X. H.; Zhang, Y. Gate-Tunable Room-Temperature Ferromagnetism in Two-Dimensional Fe3GeTe2. *Nature* **2018**, *563* (7729), 94–99. https://doi.org/10.1038/s41586-018-0626-9.

(15) Du, K.; Wang, X.; Liu, Y.; Hu, P.; Utama, M. I. B.; Gan, C. K.; Xiong, Q.; Kloc, C. Weak Van Der Waals Stacking, Wide-Range Band Gap, and Raman Study on Ultrathin Layers of Metal Phosphorus Trichalcogenides. *ACS Nano* **2016**, *10* (2), 1738–1743. https://doi.org/10.1021/acsnano.5b05927.

(16) Lee, J.-U.; Lee, S.; Ryoo, J. H.; Kang, S.; Kim, T. Y.; Kim, P.; Park, C.-H.; Park, J.-G.; Cheong, H. Ising-Type Magnetic Ordering in Atomically Thin FePS3. *Nano Lett.* **2016**, *16* (12), 7433–7438. https://doi.org/10.1021/acs.nanolett.6b03052.

(17) Otrokov, M. M.; Rusinov, I. P.; Blanco-Rey, M.; Hoffmann, M.; Vyazovskaya, A. Yu.; Eremeev, S. V.; Ernst, A.; Echenique, P. M.; Arnau, A.; Chulkov, E. V. Unique Thickness-Dependent Properties of the van Der Waals Interlayer Antiferromagnet MnBi2Te4 Films. *Phys. Rev. Lett.* **2019**, *122* (10), 107202. https://doi.org/10.1103/PhysRevLett.122.107202.

(18) Bedoya-Pinto, A.; Ji, J.-R.; Pandeya, A. K.; Gargiani, P.; Valvidares, M.; Sessi, P.; Taylor, J. M.; Radu, F.; Chang, K.; Parkin, S. S. P. Intrinsic 2D-XY Ferromagnetism in a van Der Waals Monolayer. *Science* **2021**, *374* (6567), 616–620. https://doi.org/10.1126/science.abd5146.

(19) Kim, H. H.; Yang, B.; Li, S.; Jiang, S.; Jin, C.; Tao, Z.; Nichols, G.; Sfigakis, F.; Zhong, S.; Li, C.; Tian, S.; Cory, D. G.; Miao, G.-X.; Shan, J.; Mak, K. F.; Lei, H.; Sun, K.; Zhao, L.; Tsen, A. W. Evolution of Interlayer and Intralayer Magnetism in Three Atomically Thin Chromium Trihalides. *Proc. Natl. Acad. Sci.* **2019**, *116* (23), 11131–11136. https://doi.org/10.1073/pnas.1902100116.

(20) Lee, K.; Dismukes, A. H.; Telford, E. J.; Wiscons, R. A.; Wang, J.; Xu, X.; Nuckolls, C.; Dean, C. R.; Roy, X.; Zhu, X. Magnetic Order and Symmetry in the 2D Semiconductor CrSBr. *Nano Lett.* **2021**, *21* (8), 3511–3517. https://doi.org/10.1021/acs.nanolett.1c00219.

(21) Jin, C.; Kou, L. Two-Dimensional Non-van Der Waals Magnetic Layers: Functional Materials for Potential Device Applications. *J. Phys. D: Appl. Phys.* **2021**, *54* (41), 413001. https://doi.org/10.1088/1361-6463/ac08ca.

(22) Puthirath Balan, A.; Radhakrishnan, S.; Woellner, C. F.; Sinha, S. K.; Deng, L.; Reyes, C. de los; Rao, B. M.; Paulose, M.; Neupane, R.; Apte, A.; Kochat, V.; Vajtai, R.; Harutyunyan, A. R.; Chu, C.-W.; Costin, G.; Galvao, D. S.; Martí, A. A.; van Aken, P. A.; Varghese, O. K.; Tiwary, C. S.; Malie Madom Ramaswamy Iyer, A.; Ajayan, P. M. Exfoliation of a Non-van Der Waals Material from Iron Ore Hematite. *Nat. Nanotechnol.* **2018**, *13* (7), 602–609. https://doi.org/10.1038/s41565-018-0134-y.

(23) Puthirath Balan, A.; Radhakrishnan, S.; Kumar, R.; Neupane, R.; Sinha, S. K.; Deng, L.; de los Reyes, C. A.; Apte, A.; Rao, B. M.; Paulose, M.; Vajtai, R.; Chu, C. W.; Costin, G.; Martí, A. A.; Varghese, O. K.; Singh, A. K.; Tiwary, C. S.; Anantharaman, M. R.; Ajayan, P. M. A Non-van Der Waals Two-Dimensional Material from Natural Titanium Mineral Ore Ilmenite. *Chem. Mater.* **2018**, *30* (17), 5923–5931. https://doi.org/10.1021/acs.chemmater.8b01935.

(24) Yadav, T. P.; Shirodkar, S. N.; Lertcumfu, N.; Radhakrishnan, S.; Sayed, F. N.; Malviya, K. D.; Costin, G.; Vajtai, R.; Yakobson, B. I.; Tiwary, C. S.; Ajayan, P. M. Chromiteen: A New 2D Oxide Magnetic Material from Natural Ore. *Advanced Materials Interfaces* **2018**, *5* (19), 1800549. https://doi.org/10.1002/admi.201800549.

(25) Chu, J.; Zhang, Y.; Wen, Y.; Qiao, R.; Wu, C.; He, P.; Yin, L.; Cheng, R.; Wang, F.; Wang, Z.; Xiong, J.; Li, Y.; He, J. Sub-Millimeter-Scale Growth of One-Unit-Cell-Thick Ferrimagnetic Cr2S3 Nanosheets. *Nano Lett.* **2019**, *19* (3), 2154–2161. https://doi.org/10.1021/acs.nanolett.9b00386.

(26) Zhang, Y.; Chu, J.; Yin, L.; Shifa, T. A.; Cheng, Z.; Cheng, R.; Wang, F.; Wen, Y.; Zhan, X.; Wang, Z.; He, J. Ultrathin Magnetic 2D Single-Crystal CrSe. *Adv. Mater.* **2019**, *31* (19), 1900056. https://doi.org/10.1002/adma.201900056.

(27) Wang, M.; Kang, L.; Su, J.; Zhang, L.; Dai, H.; Cheng, H.; Han, X.; Zhai, T.; Liu, Z.; Han, J. Two-Dimensional Ferromagnetism in CrTe Flakes down to Atomically Thin Layers. *Nanoscale* **2020**, *12* (31), 16427–16432. https://doi.org/10.1039/D0NR04108D.

(28) Wen, Y.; Liu, Z.; Zhang, Y.; Xia, C.; Zhai, B.; Zhang, X.; Zhai, G.; Shen, C.; He, P.; Cheng, R.; Yin, L.; Yao, Y.; Getaye Sendeku, M.; Wang, Z.; Ye, X.; Liu, C.; Jiang, C.; Shan, C.; Long, Y.; He, J. Tunable Room-Temperature Ferromagnetism in Two-Dimensional Cr2Te3. *Nano Lett.* **2020**, *20* (5), 3130–3139. https://doi.org/10.1021/acs.nanolett.9b05128.

(29) Puthirath Balan, A.; Radhakrishnan, S.; Neupane, R.; Yazdi, S.; Deng, L.; A. de los Reyes, C.; Apte, A.; B. Puthirath, A.; Rao, B. M.; Paulose, M.; Vajtai, R.; Chu, C.-W.; Martí, A. A.; Varghese, O. K.; Tiwary, C. S.; Anantharaman, M. R.; Ajayan, P. M. Magnetic Properties and Photocatalytic Applications of 2D Sheets of Nonlayered Manganese Telluride by Liquid Exfoliation. *ACS Appl. Nano Mater.* **2018**, *1* (11), 6427–6434. https://doi.org/10.1021/acsanm.8b01642.

(30) Kang, L.; Ye, C.; Zhao, X.; Zhou, X.; Hu, J.; Li, Q.; Liu, D.; Das, C. M.; Yang, J.; Hu, D.; Chen, J.; Cao, X.; Zhang, Y.; Xu, M.; Di, J.; Tian, D.; Song, P.; Kutty, G.; Zeng, Q.; Fu, Q.; Deng, Y.; Zhou, J.; Ariando, A.; Miao, F.; Hong, G.; Huang, Y.; Pennycook, S. J.; Yong, K.-T.; Ji, W.; Renshaw Wang, X.; Liu, Z. Phase-Controllable Growth of Ultrathin 2D Magnetic FeTe Crystals. *Nat. Commun.* **2020**, *11* (1), 3729. https://doi.org/10.1038/s41467-020-17253-x.

(31) Yin, J.-X.; Lian, B.; Hasan, M. Z. Topological Kagome Magnets and Superconductors. *Nature* **2022**, *612* (7941), 647–657. https://doi.org/10.1038/s41586-022-05516-0.

(32) Ghimire, N. J.; Mazin, I. I. Topology and Correlations on the Kagome Lattice. *Nat. Mater.* **2020**, *19* (2), 137–138. https://doi.org/10.1038/s41563-019-0589-8.

(33) Komędera, K.; Jasek, A. K.; Błachowski, A.; Ruebenbauer, K.; Krztoń-Maziopa, A. Magnetic Anisotropy in FeSb Studied by 57Fe Mössbauer Spectroscopy. *Journal of Magnetism and Magnetic Materials* **2016**, *399*, 221–227. https://doi.org/10.1016/j.jmmm.2015.09.076.

(34) Dong, J.; Zhang, L.; Dai, X.; Ding, F. The Epitaxy of 2D Materials Growth. *Nat Commun* **2020**, *11* (1), 5862. https://doi.org/10.1038/s41467-020-19752-3.

(35) Dong, J.; Liu, Y.; Ding, F. Mechanisms of the Epitaxial Growth of Two-Dimensional Polycrystals. *npj Comput Mater* **2022**, *8* (1), 1–11. https://doi.org/10.1038/s41524-022-00797-5.

(36) Yu, Y.; Fu, H.; She, L.; Lu, S.; Guo, Q.; Li, H.; Meng, S.; Cao, G. Fe on Sb(111): Potential Two-Dimensional Ferromagnetic Superstructures. *ACS Nano* **2017**, *11* (2), 2143–2149. https://doi.org/10.1021/acsnano.6b08347.

(37) Song, Y.-H.; Jia, Z.-Y.; Zhang, D.; Zhu, X.-Y.; Shi, Z.-Q.; Wang, H.; Zhu, L.; Yuan, Q.-Q.; Zhang, H.; Xing, D.-Y.; Li, S.-C. Observation of Coulomb Gap in the Quantum Spin Hall Candidate Single-Layer 1T'-WTe2. *Nat. Commun.* **2018**, *9* (1). https://doi.org/10.1038/s41467-018-06635-x.

(38) Ugeda, M. M.; Pulkin, A.; Tang, S.; Ryu, H.; Wu, Q.; Zhang, Y.; Wong, D.; Pedramrazi, Z.; Martín-Recio, A.; Chen, Y.; Wang, F.; Shen, Z.-X.; Mo, S.-K.; Yazyev, O. V.; Crommie, M. F. Observation of Topologically Protected States at Crystalline Phase Boundaries in Single-Layer WSe2. *Nature Communications* **2018**, *9* (1). https://doi.org/10.1038/s41467-018-05672-w.

(39) Li, Z.; Zhuang, J.; Wang, L.; Feng, H.; Gao, Q.; Xu, X.; Hao, W.; Wang, X.; Zhang, C.; Wu, K.; Dou, S. X.; Chen, L.; Hu, Z.; Du, Y. Realization of Flat Band with Possible Nontrivial Topology in Electronic Kagome Lattice. *Sci. Adv.* **2018**, *4* (11), eaau4511. https://doi.org/10.1126/sciadv.aau4511.

(40) Liu, Z. Q.; Lü, W. M.; Lim, S. L.; Qiu, X. P.; Bao, N. N.; Motapothula, M.; Yi, J. B.; Yang, M.; Dhar, S.; Venkatesan, T.; Ariando. Reversible Room-Temperature Ferromagnetism in Nb-Doped SrTiO3 Single Crystals. *Phys. Rev. B* **2013**, *87* (22), 220405. https://doi.org/10.1103/PhysRevB.87.220405.

(41) Okamoto, H. Fe-Sb (Iron-Antimony). *JPE* **1999**, *20* (2), 166–166. https://doi.org/10.1007/s11669-999-0017-x.

(42) Raghavan, V. Fe-Sb-Sn (Iron-Antimony-Tin). *JPE* **2001**, *22* (6), 669–670. https://doi.org/10.1007/s11669-001-0035-9.

(43) Yin, J.-X.; Ma, W.; Cochran, T. A.; Xu, X.; Zhang, S. S.; Tien, H.-J.; Shumiya, N.; Cheng, G.; Jiang, K.; Lian, B.; Song, Z.; Chang, G.; Belopolski, I.; Multer, D.; Litskevich, M.; Cheng, Z.-J.; Yang, X. P.; Swidler, B.; Zhou, H.; Lin, H.; Neupert, T.; Wang, Z.; Yao, N.; Chang, T.-R.; Jia, S.; Zahid Hasan, M. Quantum-Limit Chern Topological Magnetism in TbMn6Sn6. *Nature* **2020**, *583* (7817), 533–536. https://doi.org/10.1038/s41586-020-2482-7.

(44) Ye, L.; Kang, M.; Liu, J.; von Cube, F.; Wicker, C. R.; Suzuki, T.; Jozwiak, C.; Bostwick, A.; Rotenberg, E.; Bell, D. C.; Fu, L.; Comin, R.; Checkelsky, J. G. Massive Dirac Fermions in a Ferromagnetic Kagome Metal. *Nature* **2018**, *555* (7698), 638–642. https://doi.org/10.1038/nature25987.

(45) Song, C.-L.; Wang, Y.-L.; Jiang, Y.-P.; Li, Z.; Wang, L.; He, K.; Chen, X.; Ma, X.-C.; Xue, Q.-K. Molecular-Beam Epitaxy and Robust Superconductivity of Stoichiometric FeSe Crystalline Films on Bilayer Graphene. *Phys. Rev. B* **2011**, *84* (2), 020503. https://doi.org/10.1103/PhysRevB.84.020503.

(46) Xiao, W.; Kushvaha, S. S.; Wang, X. Nucleation and Growth of Aluminum on an Inert Substrate of Graphite. *J. Phys.: Condens. Matter* **2008**, *20* (22), 225002. https://doi.org/10.1088/0953-8984/20/22/225002.

(47) Perdew, J. P.; Burke, K.; Ernzerhof, M. Generalized Gradient Approximation Made Simple. *Phys. Rev. Lett.* **1996**, *77* (18), 3865–3868. https://doi.org/10.1103/PhysRevLett.77.3865.

(48) Grimme, S.; Antony, J.; Ehrlich, S.; Krieg, H. A Consistent and Accurate Ab Initio Parametrization of Density Functional Dispersion Correction (DFT-D) for the 94 Elements H-Pu. *J. Chem. Phys.* **2010**, *132* (15), 154104. https://doi.org/10.1063/1.3382344.

**TABLE OF CONTENTS**

Bilayer FeSb/SrTiO$_3$(001) films terminated with a Kagome lattice exhibit ferromagnetic order over 300 K.

**Room-temperature ferromagnetism in epitaxial bilayer FeSb/SrTiO$_3$(001) terminated with a Kagome lattice**

Huimin Zhang,[1,2]†* Qinxi Liu,[3]† Liangzi Deng,[4]† Yanjun Ma,[1] Samira Daneshmandi,[4] Cheng Cen,[1,5] Chenyu Zhang,[6] Paul M. Voyles,[6] Xue Jiang,[2,3]* Jijun Zhao,[2,3] Ching-Wu Chu,[4] Zheng Gai[7] and Lian Li[1]*

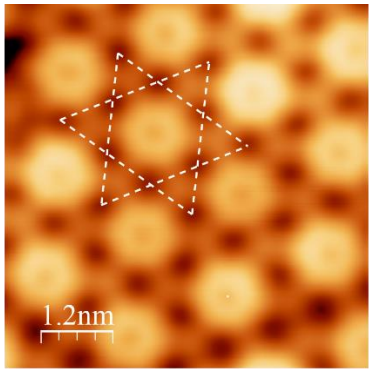 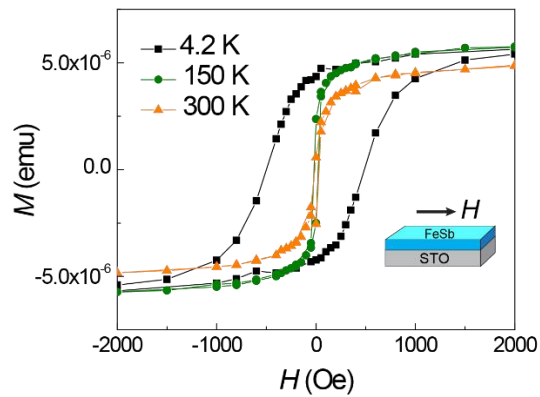